\documentclass[aps,prd,superscriptaddress,twoside,twocolumn,nofootinbib,10pt,%
showpacs,floatfix]{revtex4-1}
\usepackage[usenames]{xcolor}
\usepackage{graphicx,subfig}
\usepackage{amsmath,amssymb}
\usepackage{epstopdf}
\usepackage{bm,bbm}
\usepackage{ulem}

\renewcommand\d{\partial}
\newcommand\tr{\mathop{\mathrm{Tr}}}

\allowdisplaybreaks

\begin{document}

\title{Effects of scalar mesons in a Skyrme model with hidden local symmetry}

\author{Bing-Ran He}
\email{he@hken.phys.nagoya-u.ac.jp}
\affiliation{Department of Physics, Nagoya University, Nagoya, 464-8602, Japan}

\author{Yong-Liang Ma}
\email{yongliangma@jlu.edu.cn}
\affiliation{Center of Theoretical Physics and College of Physics, Jilin University, Changchun,
130012, China}

\author{Masayasu Harada}
\email{harada@hken.phys.nagoya-u.ac.jp}
\affiliation{Department of Physics, Nagoya University, Nagoya, 464-8602, Japan}

\date{\today}

\begin{abstract}
We study the effects of light scalar mesons on the skyrmion properties by constructing and examining a mesonic model including
pion, rho meson, and omega meson fields as well as two-quark and four-quark scalar meson fields. In our model, the physical scalar
mesons are defined as mixing states of the two- and four-quark fields. We first omit the four-quark scalar meson field from the model and find that when there is no direct coupling between the two-quark scalar meson and the vector mesons, the soliton mass is smaller and the soliton size is larger for lighter scalar mesons;
 when direct coupling is switched on, as the coupling strength increases, the soliton becomes heavy, and the radius of the baryon number density becomes large, as the repulsive force arising from the $\omega$ meson becomes strong. 
We then include the four-quark scalar meson field in the model and find that mixing between the two-quark and four-quark components of the scalar meson fields also affects the properties of the soliton. When the two-quark component of the lighter scalar meson is increased, the soliton mass decreases and the soliton size increases.
\end{abstract}

%11.30.Rd Chiral symmetries
%12.39.Dc Skyrmions
%12.39.Fe Chiral Lagrangians
%14.40.Be Light mesons (S=C=B=0)


\maketitle

%%%%%%%%%%%%%%%%%%%%%%%%%%%%%%%%%%%%%%%%%%%%%%%%%%%%%%%%%%%%%%%%%%%%%%%%%%%%%
\section{Introduction}
\label{sec:intro}

More than 50 years ago, T.~H.~R.~Skyrme proposed his pioneering idea that one may identify a certain nontrivial
topological field configuration (soliton) in mesonic theory with the baryon~\cite{Skyrme:1962vh}.
Since then, this idea has been extensively applied in
particle physics, nuclear physics, and also condensed matter physics~\cite{multifacet}.
 The originally proposed Skyrme model include only the degree of freedom of the pion.
Motivated by the essential roles of meson resonances such as the rho meson, omega meson,
and light scalar mesons in nuclear physics, researchers extended the Skyrme model to Skyrme-type models to include such resonances~\cite{Zahed:1986qz,Park:2009bb}.

Recently, especially after the appearance of holographic models of quantum chromodynamics (QCD), the effects of some resonances on the skyrmion properties were widely explored. Some key points related to
the present work are summarized as follows: (i) by dimensionally deconstructing a Yang-Mills theory in a flat
four-dimensional space in the large $N_c$ limit, it was found that as more isovector vector resonances are included,
the soliton becomes lighter~\cite{sutcliffe11};
 (ii) by using a chiral effective theory of vector mesons based on the hidden local symmetry (HLS) approach written up to
 the next-to-leading order, the omega meson that enters the chiral effective theory through the
 homogeneous Wess-Zumino term was found to provide a strong repulsive force~\cite{Ma:2012kb,Ma:2012zm};
(iii) a scalar meson that is regarded as the Nambu-Goldstone boson associated with spontaneous
breaking of the scale symmetry, i.e.,  the
dilaton, was introduced to
investigate its attractive effect~\cite{Park:2003sd,Park:2008zg,Ma:2013ela}.
It was found that the dilaton can provide an attractive force of about $100~$MeV in the model, which can reproduce chiral restoration in matter.

In this paper, we explore the effect of scalar mesons made of two quarks and four quarks on the mass and size of
the skyrmion by first constructing a chiral effective model including
the pion, the rho meson, the omega meson, and two-quark and four-quark scalar mesons.
Then, determining the parameters of the model through meson dynamics, we study the effects of  the masses and
constituents of the scalar mesons, as well as the strength of an interaction among the vector and scalar mesons, on the
mass and  size of the skyrmion.

Our findings can be summarized as follows:
\begin{enumerate}
\item When we switch off mixing between the two-quark and four-quark states in the model,
the four-quark scalar state decouples from the model. The light scalar meson is a pure two-quark state. In
such a case, we find that:
\begin{enumerate}
\item
The soliton mass and size decrease with increasing mass of the light scalar meson.

\item
The
soliton mass depends on the coupling strength between the light scalar meson and the vector mesons. The result shows that the soliton mass and the size of the baryon number density increase when
the coupling strength increases.
\end{enumerate}
\item
When we switch on mixing between the two-quark and four-quark states in the model, we find that, when the two-quark
component of the lighter scalar meson is increased, the soliton mass becomes small and its size becomes large.

\end{enumerate}

This paper is organized as follows. The next section describes how we construct
 the chiral effective model that we use in this paper and explains its typical features in relation to this work.
In sec. 3, we calculate the skyrmion properties numerically and analyze the effects of scalar
mesons on them. The last section discusses the results we obtained and presents our conclusions. In the appendixes, we give some details of the calculation.

%%%%%%%%%%%%%%%%%%%%%%%%%%%%%%%%%%%%%%%%%%%%%%%%%%%%%%%%%%%%%%%%%%%%%%%%%%%%%
\section{The model}
\label{sec:model}

We start by constructing a chiral effective model including two-quark and four-quark scalar states. Here, we
limit our discussion to  the mesons, including up and down quarks. In our model, we introduce
a two-quark field $M_{(2)}$ that is schematically written at the quark level as
\begin{equation}
{\left( M_{(2)} \right)^j}_i \ \sim \ \bar{q}_{R\,i}\,q_{L\,j} \ ,
\label{quarkM2}
\end{equation}
 where $i,j =1,2$ are the flavor indices.
This, at the hadron level, is rewritten in terms of the isosinglet $\sigma$ field and isotriplet $\pi$ fields as
\begin{eqnarray}
M_{(2)} = \frac{1}{2} \left( \sigma + i \vec{\pi} \cdot \vec{\tau} \right) \ ,
\end{eqnarray}
where $\vec{\tau}$ are the Pauli matrices. Under chiral transformation, the matrix $M_{(2)}$ transforms as
\begin{equation}
M_{(2)} \to  g_L M_{(2)} g_R^\dag\ ,
\end{equation}
where $g_{L,R} \in \mbox{SU}(2)_{L,R}$.
Furthermore, we can impose $Z_2$ symmetry as a remnant of the U(1)$_A$ transformation,
under which $q_L$ and $q_R$ transform as $q_L \to - q_L$ and $q_R \to q_R$, respectively. Therefore, the hadron field $M_{(2)}$ transforms as
\begin{align}
M_{(2)} \to- M_{(2)}\ .
\end{align}
We introduce a four-quark field that is schematically written at %in
the quark level as
\begin{equation}
\phi \ \sim\ \bar{q}_{Li}\bar{q}_{Lj}\epsilon^{ij} \, q_R^k q_R^l \epsilon_{kl} \ .
\label{eq5}
\end{equation}
In this work, we regard $\phi$ as a real field by considering only the real part of Eq.~(\ref{eq5}).
Under the $Z_2$ transformation, $\phi$ transforms as
\begin{align}
\phi \to \phi \ .
\end{align}

 On the basis of chiral symmetry and $Z_2$ symmetry, the Lagrangian is written in the form
\begin{eqnarray}
\mathcal{L} & = & \tr\left(\d_\mu M_{(2)} \d^\mu M^\dag_{(2)}\right) +
\frac{1}{2} \partial_\mu \phi \partial^\mu \phi \nonumber \\
& & {} - \left(V_0 - \bar{V}_{0}\right) - \left(V_{\rm SB} - \bar{V}_{\rm SB}\right)
\ ,\label{eq:Lagscalar}
\end{eqnarray}
where $V_0$ represents the meson potential term, and  $V_{\rm SB}$ denotes the explicit chiral symmetry breaking
term,  which has values of $\bar{V}_{0}$ and $\bar{V}_{\rm SB}$, respectively, in vacuum.  We note that, in this analysis, we include only the nonderivative interactions.
We require that the number of fields in each vertex derived from the potential  $V_0$ is less than or equal to
four, and
the number of quarks included in each vertex is
less than or equal to eight~\cite{Fariborz:2009cq}. Then, the potential is expressed as
\begin{widetext}
\begin{eqnarray}
V_0 & = & \lambda \tr\left(M_{(2)}M_{(2)}^\dag M_{(2)}M_{(2)}^\dag\right) - m_2^2 \tr\left(M_{(2)}M_{(2)}^\dag\right) +
\frac{1}{2}  m_4^2 \phi^2
\nonumber \\
& &{} + \sqrt{2} A \left( \det \left(M_{(2)} \right)
+ \det \left(M_{(2)}^\dag\right)\right) \phi \, .
\end{eqnarray}
\end{widetext}
We adopt the simplest form of the explicit chiral symmetry breaking potential $V_{\rm SB}$:
\begin{eqnarray}
V_{\rm SB} & = &{} -\frac12 f_\pi \tr \left(\chi M^\dag_{(2)} + \chi^\dag M_{(2)}\right) \ ,
\end{eqnarray}
where $\chi = 2 B {\mathcal M}$, and $B$ and ${\mathcal M}$ are a constant with dimension one and the quark mass
matrix, respectively. In this analysis, we consider an isospin-symmetric case, that is, ${\mathcal M} =\mbox{diag}( \bar{m} \,,\, \bar{m} )$, where $\bar{m} = (m_u + m_d)/2$.
We use the pion mass as an input to determine the relevant parameters, so $V_{\rm SB}$ is expressed as
\begin{equation}
V_{\rm SB} = {} -\frac{1}{2} f_\pi m_\pi^2 \tr \left(M^\dag_{(2)} + M_{(2)}\right) \ .
\end{equation}

In this paper, we introduce the rho and omega mesons based on the HLS~\cite{Bando:1987br,Harada:2003jx}.
For this purpose, we take the polar decomposition of $M_{(2)}$ as $M_{(2)}=\frac12 \xi_L^\dag \sigma \xi_R$, where $\xi_L$ and $\xi_R$, in the unitary gauge of the HLS, are
expressed by the pion fields as
\begin{equation}
\xi_R = \xi_L^\dag = e^{i \vec{\pi} \cdot \vec{\tau}/(2f_\pi) } \ .
\end{equation}
Then, we write the Lagrangian \eqref{eq:Lagscalar} as
\begin{eqnarray}
\mathcal{L} & = & \frac12 \d_\mu\sigma \d^\mu\sigma + \sigma^2\tr\left(\alpha_{\perp\mu}\alpha_\perp^\mu\right) + \frac12
\d_\mu \phi \d^\mu \phi \nonumber\\
& & {} - \left(V_0 - \bar{V}_{0}\right) - \left(V_{\rm SB} - \bar{V}_{\rm SB}\right) + \mathcal{L}_V\,,
\label{lagrangian}
\end{eqnarray}
where
\begin{equation}
\alpha_{\perp\mu} = \frac{1}{2i} \left( \partial_\mu \xi_R \cdot \xi_R^\dag - \partial_\mu \xi_L \cdot \xi_L^\dag
\right)\ ,
\end{equation}
and the newly introduced term ${\mathcal L}_V$ represents the Lagrangian for the vector meson component, which will be explicitly given later. The potentials $V_0$ and $V_{\rm SB}$ are rewritten as
\begin{eqnarray}
V_0 & = & \frac18 \lambda \sigma^4 - \frac12 m_2^2 \sigma^2 + \frac12 m_4^2 \phi^2 +\frac{1}{\sqrt{2}}A\sigma^2\phi,
\nonumber\\
V_{\rm SB} & = &{} - \frac14 m_\pi^2 f_\pi \sigma \tr\left(U+U^\dag\right),
\end{eqnarray}
where $U=\xi_L^\dag \xi_R=e^{i \vec{\pi}\cdot\vec{\tau}/f_\pi}$.

The stationary conditions of the potential of the system are obtained by taking the first derivative of the potential $V=V_0+V_{\rm SB}$ with respect to $\sigma$ and $\phi$: 
\begin{eqnarray}
\frac{\d V}{\d \sigma} & = & \frac12 \lambda \sigma^3 - m_2^2 \sigma + \sqrt{2}A\sigma \phi -m_\pi^2 f_\pi
\label{extremumsigma}\,,\\
\frac{\d V}{\d \phi} & = & m_4^2 \phi + \frac{1}{\sqrt{2}}A\sigma^2\label{extremumphi}\,.
\end{eqnarray}
After a suitable choice of the model parameters, $\sigma$ acquires its expectation value ${\sigma}_{\rm vac}$, which is identified as the pion decay constant.
Through the interaction between the two-quark and four-quark scalar states, a nonzero $A$ leads to a nonzero expectation value of $\phi$, i.e., ${\phi}_{\rm
vac}$.
The physical scalar meson fields are obtained from the fluctuations with respect to the relevant expectation values
through $\sigma = f_\pi + \tilde{\sigma}$ and $\phi = {\phi}_{\rm vac} + \tilde{\phi}$, where ${\phi}_{\rm vac} = - \frac{A f_{\pi }^2}{\sqrt{2} m_4^2}$, as given in Eq.~\eqref{para_4}.

Taking the second-order derivative, one obtains the mass matrix for $\sigma$ and $\phi$ as
\begin{eqnarray}
\left(\begin{array}{cc}
m_{\sigma}^2 & m_{\sigma\phi}^2 \\
 m_{\phi\sigma}^2 & m_{\phi}^2\\
\end{array}\right) & = & \left(
\begin{array}{cc}
\lambda f_\pi^2 + m_\pi^2 & \sqrt{2} A f_\pi \\
\sqrt{2} A f_\pi & m_4^2
\end{array}\right)\,.
\end{eqnarray}
The physical states
$f_{500}$ and $f_{1370}$ are the mixing states of $\sigma$ and $\phi$ through the rotation
\begin{eqnarray}
\left(\begin{array}{c}
f_{500}\\
f_{1370}\\
\end{array}\right) & = &
\left(\begin{array}{rr}
\cos\theta & -\sin\theta \\
\sin\theta &\cos\theta \\
\end{array}\right)
\left(\begin{array}{c}
\tilde\sigma\\
\tilde\phi
\end{array}\right)\ ,\label{mixing}
\end{eqnarray}
where $\theta$ is the mixing angle.
In Appendix~\ref{app A}, we show some relations among
 the parameters and the physical quantities that are useful for this work.
From Eq.~\eqref{mixing} one can see that when $\cos\theta \to 0$, the physical state $f_{500}$ is dominantly a four-quark
state and $f_{1370}$ is dominantly a two-quark state, whereas when $\cos\theta \to 1$, $f_{500}$ is almost a two-quark state, but $f_{1370}$ is almost
a four-quark state.

Now, we are in a position to specify the vector meson component of the Lagrangian. Here we use the following form:
\begin{equation}
{\mathcal L}_V = {\mathcal L}_{V_0} + \mathcal{L}_{\rm anom} \ ,
\label{eq:lV}
\end{equation}
where ${\mathcal L}_{V_0}$ is the Lagrangian with only intrinsic parity-even terms, whereas $\mathcal{L}_{\rm anom}$
contains the intrinsic parity-odd terms relating to the
$\mbox{U}(2)_{L} \times \mbox{U}(2)_{R}$ chiral anomaly.

Explicitly, the intrinsic parity-even Lagrangian ${\mathcal L}_{V_0}$ is expressed as
\begin{eqnarray}
\mathcal{L}_{V_0} & = & a_{\rm hls} (s_0 \sigma^2 + (1 - s_0)F^2) \tr(\hat{\alpha}_{\| \mu} \hat{\alpha}_\| ^\mu)
\nonumber\\
& & {}
- \frac{1}{2g^2}\tr(V_{\mu\nu}V^{\mu\nu})\,,
\end{eqnarray}
where $\hat{\alpha}_\parallel^\mu $ and $V_{\mu\nu}$ are defined as
\begin{eqnarray}
\hat{\alpha}_{\parallel\mu} &=& \frac{1}{2i} \left(
  \partial_\mu \xi_{\rm R} \cdot \xi_{\rm R}^\dag +
  \partial_\mu \xi_{\rm L} \cdot \xi_{\rm L}^\dag
\right) - V_\mu \,,\nonumber\\
V_{\mu\nu} &\equiv& \partial_\mu V_\nu - \partial_\nu
V_\mu - i [ V_\mu , V_\nu ] \,,
\end{eqnarray}
with
\begin{eqnarray}
V_\mu
&=&  \frac{g}{\sqrt{2}}
\left(
\begin{array}{cc}
\frac{1}{\sqrt{2}} \left( \rho_\mu^0 + \omega_\mu \right)
  & \rho_\mu^+  \\
\rho_\mu^-
  & - \frac{1}{\sqrt{2}} \left( \rho_\mu^0 + \omega_\mu \right)
\end{array}
\right)\,.
\end{eqnarray}
Here $g$ is the gauge coupling constant of the HLS, $a_{\rm hls}$ and $s_0$ are real dimensionless parameters, and $F$ is a constant of dimension one. When $\sigma$ has its vacuum expectation value (VEV), ${\sigma}_{\rm vac} = f_\pi$, the vector meson mass is expressed as $m_{V}^2=a_{\rm hls} g^2(s_0 f_\pi^2 + (1 - s_0)F^2)$, where $a_{\rm hls} g^2 s_0 f_\pi^2$ represents the mass obtained by spontaneous chiral symmetry breaking, and
$a_{\rm hls} g^2 (1-s_0)F^2$ is the chiral invariant mass.
 In this sense, $s_0$ accounts for the magnitude of the vector meson mass coming from spontaneous chiral symmetry
 breaking. In this paper, we take $F=f_\pi$, so the vector meson mass is expressed as $m_{V}^2=a_{\rm hls} g^2
 f_\pi^2$, which is the standard form given in
 the HLS~\cite{Bando:1987br,Harada:2003jx}.
Note that for ${\mathcal L}_{V_0}$, the term linear in the $\sigma$ field is excluded by $Z_2$ symmetry, and $s_0$ can be both positive and negative.

The intrinsic parity-odd terms of the
 Lagrangian $\mathcal{L}_{\rm anom}$ in Eq.~\eqref{eq:lV} read
~\cite{Harada:2003jx}
\begin{equation}
\int d^4x \mathcal{L}_{\rm anom}= \frac{N_c}{16 \pi^2} \int _{M^4} \sum_{i=1}^{3} c_i \mathcal{L}_i\,,
\end{equation}
where $M^4$ represents the four-dimensional Minkowski space, and
\begin{eqnarray}
\mathcal{L}_1 & = & i \epsilon^{\mu\nu\sigma\rho} \tr\left(\alpha_{L\mu}\alpha_{L\nu}\alpha_{L\sigma}\alpha_{R\rho}
-\alpha_{R\mu}\alpha_{R\nu}\alpha_{R\sigma}\alpha_{L\rho}\right)\,,\nonumber\\
\mathcal{L}_2 & = & i \epsilon^{\mu\nu\sigma\rho} \tr\left(\alpha_{L \mu}\alpha_{R
\nu}\alpha_{L\sigma}\alpha_{R\rho}\right)\,,\nonumber\\
\mathcal{L}_3 & = & \epsilon^{\mu\nu\sigma\rho} \tr\left[F_{V\mu\nu} \left(\alpha_{L\sigma}\alpha_{R\rho} -
\alpha_{R\sigma}\alpha_{L\rho}\right)\right]\,,
\end{eqnarray}
in terms of the 1-form and 2-form notation with $\alpha_{L}=\hat{\alpha}_{\|}-\alpha_{\perp}$,  $\alpha_{R}=
\hat{\alpha}_{\|} + \alpha_{\perp} $, and $F_V = dV - iV^2$. The low-energy constants $c_1, c_2$, and $c_3$, which are difficult to fix from fundamental QCD, are usually estimated
 phenomenologically with respect to the experimental data~\cite{Harada:2003jx}.
In this work, we focus on the effect of scalar mesons on the skyrmion properties; therefore,
we choose the parameters $c_1 = {} - c_2 = {} -2/3$ and $c_3=0$,
which provides $\omega_\mu B^\mu$~\cite{Meissner:1986js}.

%%%%%%%%%%%%%%%%%%%%%%%%%%%%%%%%%%%%%%%%%%%%%%%%%%%%%%%%%%%%%%%%%%%%%%%%%%%%%
\section{Numerical results for the skyrmion}
\label{sec:num}

In this section, we make a numerical study of the effects of scalar mesons on
 the soliton mass and also the root-mean-square (RMS) radii of
the baryon number density and energy density.
In Table.~\ref{paralist}, we summarize the parameters of the model used in this paper.
\begin{table}
\begin{tabular}{lc}
\hline\hline
\quad$ f_\pi $ \qquad & \qquad $ 92.4 $ {\rm MeV}\quad \\
\quad$ m_\pi $  \qquad & \qquad $ 139.57 $ {\rm MeV}\quad\\
\quad$ N_c $  \qquad & \qquad $ 3 $\quad\\
\quad$ c_1 + c_2 $  \qquad & \qquad $ 0 $\quad\\
\quad$ c_1 - c_2 $  \qquad & \qquad ${} - 4/3 $\quad\\
\quad$ c_3 $  \qquad & \qquad $ 0 $\quad\\
\quad$ g $  \qquad & \qquad $ 5.80 \pm 0.91 $\quad\\
\quad$ a_{\rm hls} $  \qquad & \qquad $ 2.07 \pm 0.33 $\quad\\
\hline\hline
\end{tabular}
\caption{Values of the model parameters.}
\label{paralist}
\end{table}

\subsection{The ansatz}

To study the properties of the soliton obtained from the Lagrangian \eqref{lagrangian}, we take the standard
parametrizations for the soliton configurations. Following Refs.~\cite{Skyrme:1962vh, Meissner:1986js}, we take the
ansatz for the $\pi$, $\rho$ and $\omega$ fields as
\begin{subequations}
\begin{eqnarray}
U & = & e^{i\bm{ \tau} \cdot \hat{ \bm{r}} F(r)} \,, \\
{\bm\rho} & = & \frac{G(r)}{g r}(\hat{\bm r} \times\bm{\tau} ) \,,\\
\omega_{\mu} & = & W(r) \delta_{\mu 0} \,.
\end{eqnarray}
\label{ansatz_pirhoomega}
\end{subequations}
We parametrize the scalar meson fields $\sigma$ and $\phi$ as
\begin{subequations}
\begin{eqnarray}
\sigma & = & f_\pi \left(1+\bar{\sigma}(r) \right)\,,\\
\phi & = & \phi_{\rm vac} \left(1+\bar{\phi}(r) \right)\,,
\end{eqnarray}
\label{ansatz_sigmaphi}
\end{subequations}
where $\bar{\sigma}(r)$ and $\bar{\phi}(r)$ are dimensionless functions.
Substituting Eqs.~(\ref{ansatz_pirhoomega}) and (\ref{ansatz_sigmaphi}) into 
the Lagrangian (\ref{lagrangian}), one obtains the equations of motion for the profile functions $F(r)$, $G(r)$, $W(r)$, $\bar\sigma(r)$, and $\bar\phi(r)$. The detailed
expressions are given in Appendix~\ref{app B}.
For the solutions with the
baryon number $B=1$, the wave functions $F(r)$, $G(r)$, and $W(r)$ satisfy the following boundary conditions:
\begin{eqnarray}
& & F(0) = \pi, \qquad\;\;\; F(\infty) = 0,\nonumber\\
& & G(0) = {} - 2, \qquad G(\infty) = 0,\nonumber\\
& & W\rq{}(0) = 0, \qquad\;\; W(\infty) = 0.
\label{boundary_fgw}
\end{eqnarray}
The boundary conditions for the profile functions of the scalar mesons are determined as follows: (I) as $r \to \infty$, the
VEVs of the scalar meson fields should be reproduced, so both $\bar{\sigma}(r)$ and $\bar{\phi}(r)$ must vanish,
i.e.,
\begin{eqnarray}
\bar\sigma(\infty)  =  0\ , \qquad \bar\phi(\infty) = 0\ ;
\label{boundary_sigmaphi infinity}
\end{eqnarray}
(II) the second-order derivatives
 $\bar{\sigma}^{\prime\prime}$ and $\bar{\phi}^{\prime\prime}$ should be nonsingular as $r \to 0$, in view of the last term in Eqs.~\eqref{eq:eomsigma} and \eqref{eq:eomphi}; equivalently, 
\begin{eqnarray}
\bar\sigma\rq{}(0)  =  0\ , \qquad\bar\phi\rq{}(0) = 0 \ .
\label{boundary_sigmaphi}
\end{eqnarray}

To study  the RMS radius of the baryon number density
of the system, we specify the baryon number current by introducing
the external gauge field ${\mathcal V}_\mu$ of the $U(1)$ baryon number.
Then, the baryon number current is obtained by taking a functional derivative of the total Lagrangian \eqref{lagrangian} with respect to ${\mathcal V}_\mu$.
After an explicit calculation, the baryon number density $B_0$, which is the zeroth component of the baryon number
current, is expressed as
\begin{widetext}
\begin{eqnarray}
B_0 & = & {} - \frac{2}{3 g r^2}\left\{f_{\pi }^2 g^2 r^2 a_{\text{hls}} W \left[s_0 \bar{\sigma }^2+2 s_0 \bar{\sigma
}+1\right] \right. \nonumber\\
& & \left.\qquad\qquad{} + F^\prime \left[\alpha _2-2 G \left(-\alpha _2+\alpha _3+\alpha _2 \cos F\right)+\alpha _2
\cos^2F - 2 \alpha _2 \cos F + \left(\alpha _2-\alpha _3\right) G^2\right] \right. \nonumber\\
& &\left.\qquad\qquad{} - 2 \alpha _3 \sin F G^\prime + \alpha _1 \sin^2 F
F^\prime\right\}-\frac{\sin^2F}{2\pi^2r^2}F^\prime \, ,
\label{baryon_current}
\end{eqnarray}
where $\alpha_1$, $\alpha_2$, and $\alpha_3$ relate to $c_1$, $c_2$, and $c_3$, respectively, through Eq.~\eqref{eq:c1c2c3}.
 By using the equation of motion for $W$ in Eq.~\eqref{eq:eomW}, the baryon number density is simplified as
\begin{eqnarray}
B_0&=&\frac{1}{r^2}\frac{d}{dr}\left( \frac{4 \alpha _3 \sin F (\cos F - G - 1)}{3 g}-\frac{2 r^2 W'}{3 g}+\frac{\sin (2
F)}{8 \pi ^2}-\frac{F}{4 \pi ^2}\right)\, ,
\label{baryon number density}
\end{eqnarray}
\end{widetext}
which agrees with that obtained in Ref.~\cite{Meissner:1988iv}.
 Note that, by combining Eqs.~(\ref{baryon number density}) and (\ref{boundary_fgw}), one can show that the
baryon charge is correctly normalized as $\int_0^\infty dr^3 B_0(r) = 1$.

In this analyse, we take $c_1 + c_2 =0$, $c_1 - c_2 = 3/4$, and $c_3=0$; therefore, Eq.~(\ref{baryon_current})
is reduced to
\begin{eqnarray}
B_0 & = & {} - \frac{2}{3 g r^2}f_{\pi }^2 g^2 r^2 a_{\text{hls}} W \left[s_0 \bar{\sigma }^2+2 s_0 \bar{\sigma
}+1\right]  \, .
\label{baryon_current2}
\end{eqnarray}

In our numerical calculation, in addition to the soliton mass,  we also consider
 the RMS radius of the baryon number density,
$\langle r^2\rangle ^{1/2}_B$, and the RMS radius of the energy density,
$\langle r^2\rangle ^{1/2}_E$, which are defined as follows:
\begin{eqnarray}
\langle r^2\rangle ^{1/2}_B & = & \sqrt{\int^\infty_0 d^3r r^2 B_0(r)}\,,\nonumber\\
\langle r^2\rangle ^{1/2}_E & = & \sqrt{\frac{1}{M_{\rm sol}}\int^\infty_0 d^3r r^2 M_{\rm sol}(r)}\,,
\end{eqnarray}
where $M_{\rm sol}$ is the soliton mass, and $M_{\rm sol}(r)$ is the corresponding energy density given in
Eq.~(\ref{-mr^2}).

\subsection{Effects of scalar mesons}

Now we are in a position to show how the scalar mesons affect the skyrmion properties.

First, we consider $A = 0$. In this case, the four-quark field $\phi$ decouples from the other mesons. The
Lagrangian~\eqref{lagrangian} is reduced to
\begin{eqnarray}
\mathcal{L} & = & \frac12 \d_\mu\sigma \d^\mu\sigma + \sigma^2\tr(\alpha_{\perp\mu}\alpha_\perp^\mu) \nonumber\\
& & {} - (V_\sigma - \bar{V}_{\sigma}) - (V_{\rm SB} - \bar{V}_{\rm SB}) + \mathcal{L}_V\,, \label{lagrangian_2}
\end{eqnarray}
where $V_\sigma=\frac18 \lambda \sigma^4 - \frac12 m_2^2 \sigma^2$.
The parameters $\lambda$ and $m_2^2$ are related to the masses of $\sigma$ and $\pi$ through
\begin{eqnarray}
m_\sigma^2 &=&{} - m_2^2 + \frac{3}{2} \lambda f_\pi^2 \ , \nonumber\\
m_\pi^2 &=&{} - m_2^2 + \frac{1}{2} \lambda f_\pi^2 \ .
\end{eqnarray}
Here we take the empirical value of $m_\pi$ as an input and regard $m_\sigma$ as a parameter of the model.
In addition, $s_0$ in ${\mathcal L}_V$ is an unfixed parameter. In the following, we study the relation between these
two parameters and the skyrmion properties.

\paragraph{ The effects of $m_\sigma$ on the skyrmion properties.}

We start our discussion from the standard HLS Lagrangian~\cite{Harada:2003jx}, which is obtained by taking $s_0 = 0$ in this analysis.

\begin{figure}[htb]
\centering
\includegraphics[scale=0.45]{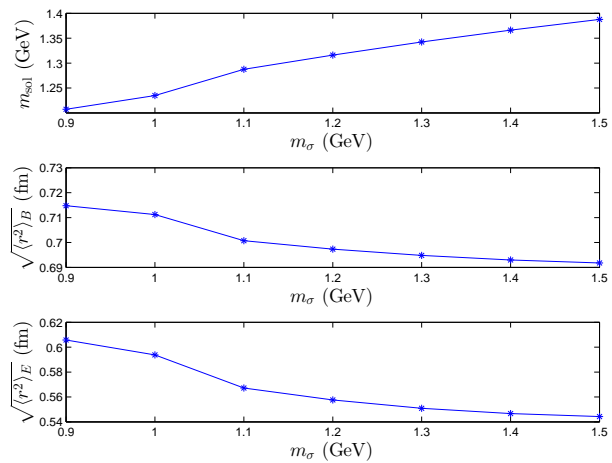}
\caption{$m_\sigma$ dependence of the soliton mass and radii with $A = s_0 = 0$. }
\label{pi_rho_omega_sigma_00}
\end{figure}

\begin{figure}[htb]
\centering
\includegraphics[scale=0.45]{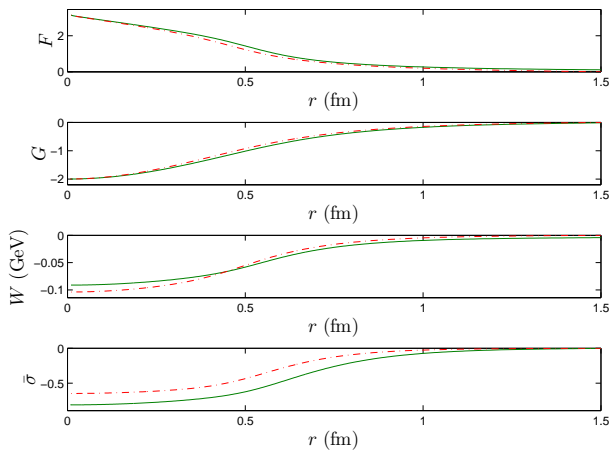}
\caption{$m_\sigma$ dependence of the profile functions with $A = s_0 = 0$ for $m_\sigma=1$\,GeV (green solid line) and
$m_\sigma=1.4$\,GeV (red dash-dotted line).
}
\label{profile_sigma}
\end{figure}

We show the $m_\sigma$ dependence of the soliton mass $m_{\rm sol}$ and the radii $\sqrt{\langle r^2\rangle _B}$ and $\sqrt{\langle r^2\rangle _E}$ in Fig.~\ref{pi_rho_omega_sigma_00} and the
profile functions in Fig.~\ref{profile_sigma}. From Fig.~\ref{pi_rho_omega_sigma_00} we find that, increasing 
$m_\sigma$, the soliton mass $m_{\rm sol}$ increases. This tendency can be understood from the profile function of 
$\bar{\sigma}$ in Fig.~\ref{profile_sigma};
when the $\sigma$  is heavy, the magnitude of $\bar{\sigma}$ is small. Because $\bar{\sigma} < 0$, as $m_\sigma$ increases, the magnitude of $\sigma=f_\pi(1+\bar{\sigma})$ becomes large; consequently, 
the contribution of $\sigma^2\tr(\alpha_{\perp\mu}\alpha_\perp^\mu)$ to the skyrmion mass becomes
large. Physically, for a heavy $\sigma$, the attractive force provided by the $\sigma$ is small; as a result, the soliton
mass is large.

In addition, Fig.~\ref{pi_rho_omega_sigma_00} tells us that when $m_\sigma$ is large, the radii $\sqrt{\langle r^2\rangle
_E}$ and $\sqrt{\langle r^2\rangle _B}$ are small. This result can be understood from the profile functions plotted in
Fig.~\ref{profile_sigma}; when $m_\sigma$ is large, the profile for $\bar{\sigma}$ becomes narrow.
 Generally, the $\sigma$ supplies the attractive force, whereas the $\omega$ meson supplies the repulsive force. A narrow $\sigma$ needs a narrow $\omega$ to balance the attractive and repulsive forces. The shapes of $\sigma$ and $\omega$ dominate the shape of the energy distribution. As a result, the narrow $\sigma$ and $\omega$ make
the radius $\sqrt{\langle r^2\rangle _E}$ small.
As for the radius of the baryon number density, from expression Eq.~(\ref{baryon_current2}) with $s_0=0$, one can easily
see that a narrow $W$ gives a small radius $\sqrt{\langle r^2\rangle _B}$.
To summarize, when $m_\sigma$ is large, the effective range of the attractive and repulsive forces supplied by the $\sigma$ and $\omega$, respectively, are small; i.e., the soliton size is small.

\paragraph{ The effects of $s_0$ on the skyrmion properties.}

\begin{figure}[htb]
\begin{center}
\includegraphics[scale=0.45]{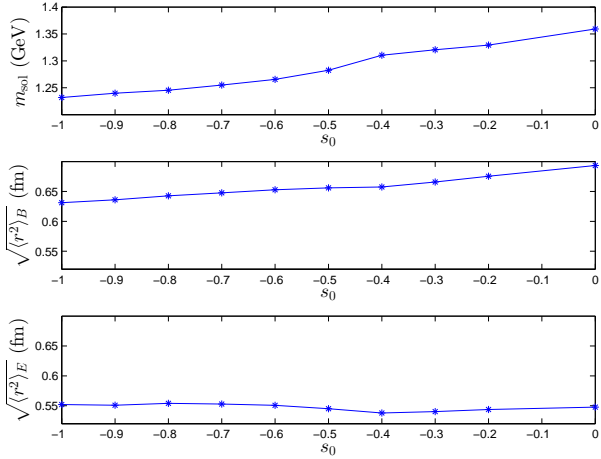}
\caption{%The parameter
$s_0$ dependence of the skyrmion mass and radii for $m_\sigma = 1.37$ GeV.
}
\label{pi_rho_omega_sigma_phi_s0}
\end{center}
\end{figure}

The parameter $s_0$ controls the chiral invariant vector meson mass. Here, we show how $s_0$ influences  the properties of the skyrmion  by taking a typical value of the scalar meson mass, $m_\sigma = 1.37$~GeV. Our numerical results are plotted in Fig.~\ref{pi_rho_omega_sigma_phi_s0}.

From Fig.~\ref{pi_rho_omega_sigma_phi_s0}, we see that, when the value of $s_0$ is increased, $m_{\rm sol}$ and $\sqrt{\langle r^2\rangle _B}$ become large. According to the equations of motion for the
vector mesons in Eq.~\eqref{eq:eomW}, $ a_{\rm hls} g_\omega^2 \left( f_\pi^2 (1-s_0) + s_0
\sigma^2(r)\right)$ acts as the effective mass of the vector mesons inside the soliton. Because $\sigma(r) < f_\pi$, 
as we stated above, the effective mass is smaller than the meson mass $m_\omega = \sqrt{a_{\rm hls}} g_\omega f_\pi$ in vacuum. This has two effects:
 on the one hand, the effective strength of the
repulsive force mediated by the $\omega$ meson is strong for large $s_0$,
which produces a large soliton mass, as the vector meson dominates the energy; on the other hand, the effective range of the repulsive force supplied by $\omega$ is long for large $s_0$, which makes the RMS radius of the baryon number density $\sqrt{\langle r^2\rangle _B}$ large, as the $\omega$ meson acts as the gauge boson of the $U(1)_{\rm V}$ baryon number symmetry. The relation between $s_0$ and $\sqrt{\langle r^2\rangle _B}$ needs a more precise explanation; because the baryon number density  is proportional to the effect of the $\omega$ meson, as can be seen in
 Eq.~\eqref{baryon_current2},
 the RMS radius of the baryon number density  $\sqrt{\langle r^2\rangle _B}$
is large for large $s_0$.

\begin{figure}[htb]
\begin{center}
\includegraphics[scale=0.45]{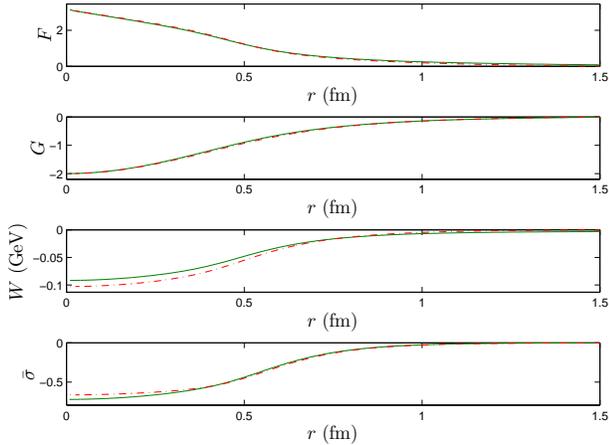}
\caption{Profile functions with $s_0 = {} -0.5$ (green line) and $s_0=0$ (red dotted line).}
\label{profile_s0}
\end{center}
\end{figure}

The situation for the RMS radius of the energy density,
$\sqrt{\langle r^2\rangle _E}$, is quite complicated. In this analysis, the numerical error is about $1$-$3\%$, so
 the third graph in Fig.~\ref{profile_s0} implies that $\sqrt{\langle r^2\rangle _E}$ is rather stable against changes in $s_0$. This implies that the contribution of $\omega$ has the opposite direction to that of $\sigma$, so they
 cancel each other.

\begin{figure}[htb]
\begin{center}
\includegraphics[scale=0.45]{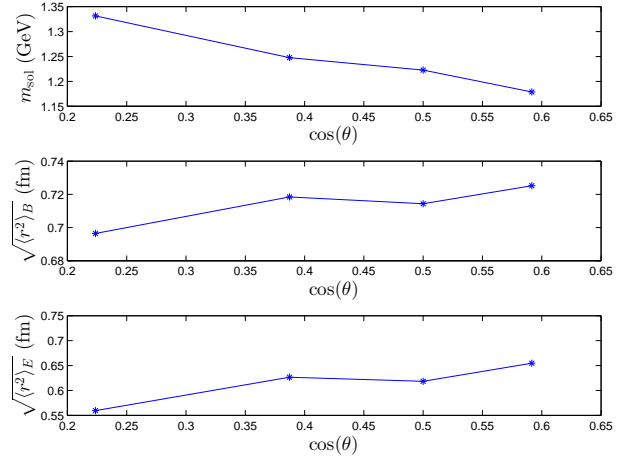}
\caption{Dependence of skyrmion properties on the mixing angle between two scalar mesons for
$s_0=0$.
}
\label{pi_rho_omega_sigma_phi_00}
\end{center}
\end{figure}

\begin{figure}[htb]
\begin{center}
\includegraphics[scale=0.45]{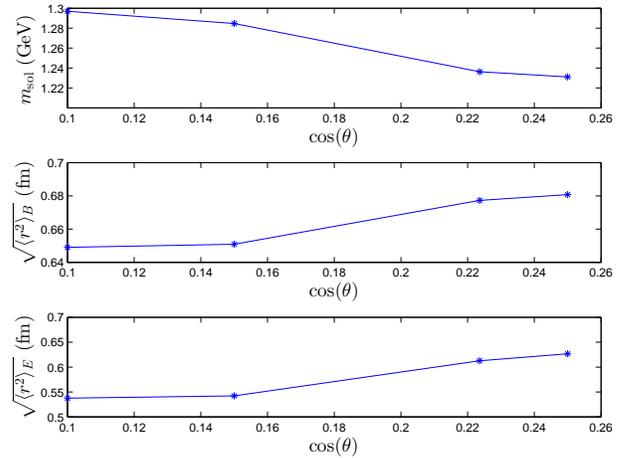}
\caption{Dependence of skyrmion properties on the mixing angle between two scalar mesons for $s_0={}-0.5$.
}
\label{pi_rho_omega_sigma_phi_05}
\end{center}
\end{figure}

Next, we switch on the mixing between the two scalar mesons by setting $A \neq 0$ and then study how the mixing angle
$\theta$ affects the skyrmion properties. In our calculation, we take two fixed values of $s_0$, i.e., $s_0 = 0$ and
$s_0 ={} - 0.5$, to see the dependence of the mass and radii of the skyrmion on the mixing angle
$\theta$ defined in Eq.~\eqref{mixing}.
Our results are shown in Figs.~\ref{pi_rho_omega_sigma_phi_00} and \ref{pi_rho_omega_sigma_phi_05}.
Note that we cannot find any stable solution in the regions of $\cos\theta > 0.6$ for $s_0 =0$ and
$\cos\theta > 0.25$ for $s_0 ={} -0.5$. In other words, the maximal value of $\cos\theta$ is small
for small $s_0$. This could be understood as follows. The definition of the mixing angle in Eq.~\eqref{mixing} implies that,
for a large value of $\cos\theta$, $f_{500}$ includes a large amount of the two-quark component $\tilde\sigma$, which generates a strong attractive force.
Because the total repulsive force does not change, the attractive force can exceed the total repulsive force, and the soliton collapses. A large value of $\cos{\theta}$ corresponds to a strong attractive force; therefore, the soliton collapses at a certain value of $\cos\theta$. On the other hand, a decrease in $s_0$ means that the repulsive force mediated by the $\omega$ meson becomes weak, as we stated above. Because the soliton is stabilized by the balance between the attractive and repulsive forces, the maximal value of $\cos\theta$ is small for small $s_0$.

Note that the decreasing and increasing tendencies in Figs.~\ref{pi_rho_omega_sigma_phi_00} and
\ref{pi_rho_omega_sigma_phi_05} are the same, and they are actually consistent with those in
Fig.~\ref{pi_rho_omega_sigma_00}. As we just stated, increasing $\cos\theta$ increases the magnitude of the two-quark component  included in $f_{500}$. Thus, the
large $\cos\theta$ in Figs.~\ref{pi_rho_omega_sigma_phi_00} and \ref{pi_rho_omega_sigma_phi_05}
 corresponds to the small mass in Fig.~\ref{pi_rho_omega_sigma_00}.

%%%%%%%%%%%%%%%%%%%%%%%%%%%%%%%%%%%%%%%%%%%%%%%%%%%%%%%%%%%%%%%%%%%%%%%%%%%%%
\section{Conclusions and discussions}
\label{sec:dis}

In this work, we first investigated the effect of a two-quark scalar meson on the skyrmion properties when mixing between the two-quark and four-quark states of scalar mesons is switched off. We found that as the mass of the 
two-quark scalar meson decreases, the soliton mass decreases and the RMS radii of the baryon number density, $\sqrt{\langle r^2\rangle _B}$, and the energy density, $\sqrt{\langle r^2\rangle _E}$, increase.  
We next studied the effect on the skyrmion properties of coupling between the scalar meson and vector mesons. It is found that as coupling becomes larger, the soliton mass and radii increase. We finally explored the skyrmion properties by considering the mixing structure of the two scalar mesons. 
We showed that, when the percentage of the two-quark component in the lighter scalar meson is increased, the soliton mass becomes small, and the RMS radii ($\sqrt{\langle r^2\rangle _B}$ and $\sqrt{\langle r^2\rangle _E}$) become large.

Our numerical results show that, even though the two-quark and four-quark scalar mesons are both included in the
model, the soliton mass obtained here is still about $300$~MeV larger than that 
needed to reproduce the empirical values of the nucleon mass and $\Delta$-$N$ mass splitting.
Thus, to reproduce the baryon masses in the real world, the present model should be extended.
There are several possible extensions, 
for example, inclusion of the effect of the dilaton as the Nambu-Goldstone boson associated with the scale symmetry breaking of QCD; the effect of the $\mathcal{O}(p^4)$ terms of the HLS, which account for the $\pi$-$\rho$ interaction; and/or the effect of 
all the homogeneous Wess-Zumino terms of the HLS, which increase the number of $\pi$-$\rho$ and/or $\pi$-$\rho$-$\omega$ terms. We leave these possibilities as  future projects. 
In addition, even though the present model cannot reproduce reality,
it still merits application to dense matter systems to study the effect of the scalar mesons on the phase
structure of nuclear matter. We will report this progress elsewhere.

%%%%%%%%%%%%%%%%%%%%%%%%%%%%%%%%%%%%%%%%%%%%%%%%%%%%%%%%%%%%%%%%%%%%%%%%%%%%%%%%%%%%%%%%%%%%%%%%%%%%%

\acknowledgments

M.~H. was supported in part by the JSPS Grant-in-Aid for Scientific Research (S) No.~22224003 and (c) No.~24540266.
And the work of Y.-L.~M. was supported in part by National Science Foundation of China (NSFC) under
Grant No.~11475071 and the Seeds Funding of Jilin University.

%%%%%%%%%%%%%%%%%%%%%%%%%%%%%%%%%%%%%%%%%%%%%%%%%%%%%%%%%%%%%%%%%%%%%%%%%%%%%%%%%%%%%%%%%%%%%%%%%%%%%

\begin{appendix}

%\begin{widetext}

\section{The mixing matrix}
\label{app A}
The masses of $f_{500}$ and $f_{1370}$ are
\begin{widetext}
\begin{eqnarray}
m_{f_{500}}^2 & = & \frac{1}{2} \left({} - \sqrt{2 f_{\pi }^2 \left(4 A^2 - \lambda  m_4^2 + \lambda  m_{\pi }^2\right) +
f_{\pi }^4 \lambda ^2 + \left(m_4^2 - m_{\pi }^2\right){}^2} + f_{\pi }^2 \lambda + m_4^2 + m_{\pi
}^2\right)\label{mf500}\,,\\
m_{f_{1370}}^2 & = & \frac{1}{2} \left(\sqrt{2 f_{\pi }^2 \left(4 A^2 - \lambda  m_4^2 + \lambda  m_{\pi }^2\right) +
f_{\pi }^4 \lambda ^2 + \left(m_4^2 - m_{\pi }^2\right){}^2} + f_{\pi }^2 \lambda + m_4^2 + m_{\pi
}^2\right)\label{mf1370}\,.
\end{eqnarray}
\end{widetext}
The mixing angle $\theta$ can be obtained as
\begin{equation}
\theta=\arctan\sqrt{\frac{m_{f_{1370}}^2-m_4^2}{m_4^2-m_{f_{500}}^2}}=\arccos\sqrt{\frac{m_4^2-m_{f_{500}}^2}{m_{f_{1370}}^2-m_{f_{500}}^2}}\label{theta}\,.
\end{equation}
 From Eqs. (\ref{extremumsigma}), (\ref{extremumphi}), (\ref{mf500}) and (\ref{mf1370}), one gets
\begin{eqnarray}
A^2 & = & \frac{ (m_{f_{1370}}^2-m_4^2) (m_4^2-m_{f_{500}}^2) }{2 f_{\pi }^2}, \nonumber\\
\lambda & = & \frac{m_{f_{500}}^2+m_{f_{1370}}^2-m_4^2-m_{\pi }^2}{f_{\pi }^2}, \nonumber\\
{\phi}_{\rm vac} & = &{} - \frac{A f_{\pi }^2}{\sqrt{2} m_4^2}, \nonumber\\
m_2^2 & = & \frac{1}{2} \left(\frac{m_{f_{500}}^2 m_{f_{1370}}^2}{m_4^2}-3 m_{\pi }^2\right),\label{para_4}
\end{eqnarray}
with $m_{f_{500}}^2<m_4^2<m_{f_{1370}}^2$.

\begin{widetext}

\section{Soliton mass and the equations of motion for the profile functions $F(r), G(r), W(r), \bar{\sigma}(r)$ and
$\bar{\phi}(r)$}
\label{app B}

Substituting the ansatz defined in Eqs.~\eqref{ansatz_pirhoomega} and \eqref{ansatz_sigmaphi} to %and
 the effective Lagrangian \eqref{lagrangian}, we obtain the following expression for the soliton mass:% as% is expressed as
\begin{eqnarray}
M_{\rm sol} & = & 4\pi \int_0^\infty dr r^2 M_{\rm sol}(r)
\nonumber
\\
& = &{} - 4\pi\int_0^\infty dr \Big\{ \frac{1}{8} r^2 \Big\{4 f_{\pi }^2 g^2 s_0 a_{\text{hls}} W^2 \bar{\sigma }
(\bar{\sigma } + 2)-4 f_{\pi }^2 \left(\bar{\sigma }+1\right)^2 F'^2\nonumber\\
&&+f_{\pi }^2 \Big[\frac{2 \left(m_4^2-m_{f_{500}}^2\right) \left(m_{f_{1370}}^2-m_4^2\right) \left(\bar{\sigma
}+1\right)^2 \left(\bar{\phi }+1\right)}{m_4^2} -\left(m_{f_{500}}^2+m_{f_{1370}}^2-m_4^2-m_{\pi }^2\right)
\left(\bar{\sigma }+1\right)^4\nonumber\\
&&+\frac{2 \left(m_{f_{500}}^2 m_{f_{1370}}^2-3 m_4^2 m_{\pi }^2\right) \left(\bar{\sigma
}+1\right)^2}{m_4^2}+\frac{\left(m_4^2-m_{f_{500}}^2\right) \left(m_4^2-m_{f_{1370}}^2\right) \left(\bar{\phi
}+1\right)^2}{m_4^2}+8 m_{\pi }^2 \left(\bar{\sigma }+1\right) \cos (F)\Big]\nonumber\\
&&+\frac{f_{\pi }^2 \left(m_4^2-m_{f_{500}}^2\right) \left(m_4^2-m_{f_{1370}}^2\right) \bar{\phi }'^2}{m_4^4}-4 f_{\pi
}^2 \bar{\sigma }'^2+4 f_{\pi }^2 g^2 a_{\text{hls}} W^2-\frac{f_{\pi }^2 \left(m_{f_{500}}^2 m_{f_{1370}}^2+3 m_4^2
m_{\pi }^2\right)}{m_4^2}+4 W'^2\Big\}\nonumber\\
&&-\frac{1}{2 g^2}\Big\{8 f_{\pi }^2 g^2 a_{\text{hls}} G \sin ^2\left(\frac{F}{2}\right) \left(s_0 \bar{\sigma }^2+2 s_0
\bar{\sigma }+1\right)-4 f_{\pi }^2 g^2 \bar{\sigma }^2 \sin ^2\left(\frac{F}{2}\right) \left(\left(s_0
a_{\text{hls}}-1\right) \cos (F)-s_0 a_{\text{hls}}-1\right)\nonumber\\
&&-8 f_{\pi }^2 g^2 \bar{\sigma } \sin ^2\left(\frac{F}{2}\right) \left(\left(s_0 a_{\text{hls}}-1\right) \cos (F) -s_0
a_{\text{hls}} -1\right)+2 f_{\pi }^2 g^2 a_{\text{hls}} G^2 \left(s_0 \bar{\sigma }^2+2 s_0 \bar{\sigma
}+1\right)\nonumber\\
&&+8 f_{\pi }^2 g^2 a_{\text{hls}} \sin ^4\left(\frac{F}{2}\right) -f_{\pi }^2 g^2 \cos (2 F)+f_{\pi }^2 g^2+2
G'^2\Big\}\nonumber\\
&&-\alpha _3 \left(2 G \left(W F'-\sin (F) W'\right)+G^2 W F'+2 \sin (F) \left((\cos (F)-1) W'+W
G'\right)\right)\nonumber\\
&&+\alpha _2 W F' (-\cos (F)+G+1)^2+\alpha _1 W F' \sin ^2(F)-\frac{G^2 (G+2)^2}{2 g^2 r^2}\Big\}\, ,
\label{-mr^2}
\end{eqnarray}
 with
\begin{eqnarray}
\alpha_1=\frac{3 g N_c}{16 \pi^2}(c_1 - c_2)\,,
\alpha_2=\frac{g N_c}{16 \pi^2}(c_1 + c_2)\,,
\alpha_3=\frac{ g N_c}{16 \pi^2}c_3\,.\label{eq:c1c2c3}
\end{eqnarray}
The equations of motion for $F(r), G(r), W(r), \bar{\sigma}(r)$ and $\bar{\phi}(r)$ are expressed as
\begin{eqnarray}
F^{\prime\prime} & = & \frac{1}{f_ {\pi }^2 r^2 (1+\bar{\sigma })^2} \nonumber\\
& & {} \times \Big\{2 G \left( f_ {\pi }^2 s_0 a_ {\text{hls}} \bar{\sigma }^2 \sin F + 2 f_ {\pi }^2 s_0 a_ {\text{hls}}
\bar{\sigma } \sin F + f_ {\pi }^2 a_ {\text{hls}} \sin F \right. \nonumber\\
& &\left. \qquad\qquad {} - \alpha _ 2 \cos F W'-\alpha _ 3 \cos F W'+\alpha _ 2 W G'-\alpha _ 3 W G'+\alpha _ 2
W'-\alpha _ 3 W' \right)\nonumber\\
& &\qquad{} - f_ {\pi }^2 \bar{\sigma } \sin F \left(4 s_0 a_ {\text{hls}} \cos F - 4 s_0 a_{\text{hls}}-4 \cos F -r^2m_
{\pi }^2 \right) -2 f_ {\pi }^2 \bar{\sigma }^2 \sin F \left(s_0 a_ {\text{hls}} \cos F - s_0 a_{\text{hls}}-\cos
F\right)\nonumber\\
& &\qquad{} - 2 f_ {\pi }^2 r^2 F' \bar{\sigma }' - 2 f_ {\pi }^2 r \bar{\sigma }^2 F' - 2 f_ {\pi }^2 r \bar{\sigma } F'
\left(r \bar{\sigma }' + 2\right) + 2 f_ {\pi }^2 a_ {\text{hls}} \sin F - f_ {\pi }^2 a_ {\text{hls}} \sin (2
F)\nonumber\\
& &\qquad{} - 2 f_ {\pi }^2 r F' + f_ {\pi }^2 m_ {\pi }^2 r^2 \sin F + f_ {\pi }^2 \sin (2 F) - 2 \alpha_2 W  \cos F G'
+ 2 \alpha _ 3 W  \cos F G'\nonumber\\
& &\qquad{} - \frac{1}{2} \alpha _ 1 \cos (2 F) W' - 2 \alpha_2 \cos F W' + \frac{1}{2} \alpha_2 \cos (2 F) W'-2 \alpha_3
\cos F W' + 2 \alpha_3 \cos (2 F) W' + 2 \alpha_2 W G'\nonumber\\
& &\qquad{} -2 \alpha _ 3 W G' +\left(\alpha _ 2-\alpha _ 3\right) G ^2 W'+\frac{1}{2} \alpha _ 1 W'+\frac{3}{2} \alpha _
2 W'\Big\}\,,\\
%%%%%%%%%%%%%%%%%%%%%%%%%%%%%%%%%%%%%%%%%%%%%%%%%%%%%%%%%%%%%%%%%%%%%%%%%%%%%%%%%%%%%%%%%%%%%%%%%%%%%%%%%%%
G''& = &{} - f_ {\pi }^2 g^2 a_ {\text{hls}} \left(s_0 \bar{\sigma }^2+2 s_0 \bar{\sigma }+1\right) (\cos F-G-1) -
\alpha_3 g^2 \left(W F' (\cos F - G - 1) + 2 \sin F W'\right)\nonumber\\
& &{} + \alpha_2 g^2 W F' (\cos F - G - 1)+\frac{G  \left(G^2 + 3 G + 2\right)}{r^2}\,,
\label{eq:eomG}\\
%%%%%%%%%%%%%%%%%%%%%%%%%%%%%%%%%%%%%%%%%%%%%%%%%%%%%%%%%%%%%%%%%%%%%%%%%%%%%%%%%%%%%%%%%%%%%%%%%%%%%%%%%%%
W'' & = & f_ {\pi }^2 g^2 a_{\text{hls}} W  \left(s_0 \bar{\sigma }^2+2 s_0 \bar{\sigma }+1\right) -\frac{2
W'}{r}\nonumber\\
& &{} + \frac{1}{r^2}\Big\{-\alpha_3 \left(F' \left(2 G (\cos F + 1) + 2 (\cos F - \cos (2 F))+G ^2\right)+4 \sin F G'
\right)\nonumber\\
& &\qquad\quad{} + \alpha _ 2 F' (-\cos F + G + 1)^2 + 2 \alpha_1 F' \sin^2\left(\frac{F}{2}\right) (\cos F + 1)\Big\}\,,
\label{eq:eomW}\\
%%%%%%%%%%%%%%%%%%%%%%%%%%%%%%%%%%%%%%%%%%%%%%%%%%%%%%%%%%%%%%%%%%%%%%%%%%%%%%%%%%%%%%%%%%%%%%%%%%%%%%%%%%%
\bar{\sigma }'' & = & \frac{1}{2 m_ 4^2}\Big\{m_ 4^2 (2 \bar{\sigma } \left(-g^2 s_0 a_{\text{hls}} W ^2+m_
{f_{500}}^2+m_ {f_{1370}}^2+F'^2-m_ 4^2\right)+\left(m_ {f_{500}}^2+m_ {f_{1370}}^2-m_ 4^2-m_ {\pi }^2\right) \bar{\sigma
}^3\nonumber\\
& &\qquad\;{} + 3 \left(m_ {f_{500}}^2+m_ {f_{1370}}^2-m_ 4^2-m_ {\pi }^2\right) \bar{\sigma }^2 +2 \left[{} - g^2 s_0
a_{\text{hls}} W ^2+F'^2 - m_ {\pi }^2 (\cos F - 1 )\right] \nonumber\\
& &\qquad\;{} -\left(m_ 4^2-m_ {f_{500}}^2\right) \left(m_ {f_{1370}}^2-m_ 4^2\right) \left(\bar{\sigma }+1\right)
\bar{\phi }\Big\}\nonumber\\
& &{} - \frac{\left(\bar{\sigma }+1\right) \left(-8 s_0 a_{\text{hls}} G  \sin ^2\left(\frac{F}{2}\right)-8 s_0 a_
{\text{hls}} \sin ^4\left(\frac{F}{2}\right)-2 s_0 a_{\text{hls}} G ^2+\cos (2 F)-1\right)}{r^2}-\frac{2 \bar{\sigma
}'}{r}\,, \label{eq:eomsigma}\\
%%%%%%%%%%%%%%%%%%%%%%%%%%%%%%%%%%%%%%%%%%%%%%%%%%%%%%%%%%%%%%%%%%%%%%%%%%%%%%%%%%%%%%%%%%%%%%%%%%%%%%%%%%%
\bar{\phi}'' & = & m_ 4^2 \left(\bar{\phi }-\bar{\sigma } \left(\bar{\sigma }+2\right)\right)-\frac{2 \bar{\phi
}'}{r}\,.\label{eq:eomphi}
\end{eqnarray}

\end{widetext}

\end{appendix}

\end{document}